\documentclass{desyproc}

\begin{document}
%------------------------------------
\title{Latest Results of the OSQAR Photon Regeneration Experiment for Axion-Like Particle Search}

%for single authors the superscripts are optional
%\author{{\slshape Axel Lindner$^1$, Konstantin Zioutas$^{2}$}\\[1ex]
%$^1$Deutsches Elektronen-Synchrotron (DESY), Hamburg, Germany\\
%$^2$University of Patras, Patras, Greece, and CERN, Geneva, Switzerland}

\author{{\slshape Rafik Ballou$^{1,2}$, Guy Deferne$^{3}$, Lionel Duvillaret$^{4}$, Michael Finger, Jr.$^{5}$, Miroslav Finger$^{5}$, Lucie Flekova$^{5}$, Jan Hosek$^{6}$, Tomas Husek$^{5}$, Vladimir Jary$^{6}$, Remy Jost$^{7,8}$, Miroslav Kral$^{6}$, Stepan Kunc$^{9}$, Karolina Macuchova$^{6}$, Krzysztof A. Meissner$^{10}$, J\'er\^ome Morville$^{11,12}$, Pierre Pugnat$^{13,14}$, Daniele Romanini$^{7,8}$, Matthias Schott$^{15}$, Andrzej Siemko$^{3}$, Miloslav Slunecka$^{5}$, Miroslav Sulc$^{9}$, Guy Vitrant$^{4}$, Christoph Weinsheimer$^{15}$, Josef Zicha$^{6}$}\\[1ex]
$^1$CNRS, Institut N\'eel, F-38042 Grenoble, France\\
$^2$Universit\'e Grenoble Alpes, Institut N\'eel, F-38042 Grenoble, France\\
$^3$CERN, CH-1211 Geneva-23, Switzerland\\
$^4$Grenoble INP - Minatec \& CNRS, IMEP-LAHC, F-38016 Grenoble, France\\
$^5$Charles University, Faculty of Mathematics and Physics, Prague, Czech Republic\\
$^6$Czech Technical University, Prague, Czech Republic\\
$^7$Universit\'e Grenoble Alpes, LIPhy, F-38000 Grenoble, France\\
$^8$CNRS, LIPhy, F-38000 Grenoble, France\\
$^9$Technical University of Liberec, 46117 Liberec, Czech Republic\\
$^{10}$University of Warsaw, Institute of Theoretical Physics, 00-681 Warsaw, Poland\\
$^{11}$Universit\'e Claude Bernard Lyon-1, Institut Lumi\`ere Mati\`ere, F-69622 Villeurbanne, France\\
$^{12}$CNRS, Institut Lumi\`ere Mati\`ere, F-69622 Villeurbanne, France\\
$^{13}$CNRS, LNCMI, F-38042 Grenoble, France\\
$^{14}$Universit\'e Grenoble Alpes, LNCMI, F-38042 Grenoble, France\\
$^{15}$University of Mainz, Institute of Physics, 55128 Mainz, Germany\\}

% if the proceedings are available online (e.g. at Indico)
% please enter the contribution ID or file_name below for the DOI
%\contribID{32}
\contribID{familyname\_firstname}

% TO THE CONFERENCE EDITORS: 
% please update the following information      
% before sending the template to the authors
\confID{300768}  % if the conference is on Indico uncomment this line
\desyproc{DESY-PROC-2014-03}
\acronym{Patras 2014} % if you want the Acronym in the page footer uncomment this line
\doi  % if there is an online version we will register DOIs

\maketitle

\begin{abstract}
The OSQAR photon regeneration experiment searches for pseudoscalar and scalar axion-like particles by the method of ``Light Shining Through a Wall'', based on the assumption that these weakly interacting sub-eV particles couple to two photons to give rise to quantum oscillations with optical photons in strong magnetic field. No excess of events has been observed, which constrains the di-photon coupling strength of both pseudoscalar and scalar particles down to $5.7 \cdot 10^{-8}$ GeV$^{-1}$ in the massless limit. This result is the most stringent constraint on the di-photon coupling strength ever achieved in laboratory experiments.
\end{abstract}

\section{Introduction}
\label{sec:introduction}

Embedding the Standard Model (SM) of particle physics into more general unified theories often results in postulating new elementary particles in unexplored parameter space. A number of weakly interacting sub-eV particles (WISPs) are thus predicted besides the weakly interacting massive particles (WIMPs). The most prominent example of WISPs is the axion~\cite{Weinberg1978}, first anticipated from the breaking at the quantum level of an additional ${U(1)}_{PQ}$ global symmetry postulated to provide a natural solution of the strong CP problem~\cite{Peccei1977}. This light spin-zero particle is one of the basic outputs of the string theory, where a number of axions and axion-like particles (ALPs), either pseudoscalar or scalar, naturally emerge~\cite{Svrek2006}. It moreover constitutes a non-supersymmetric candidate of the so far unobserved dark matter in the universe~\cite{Bradlay2003}. 

The detection of WISPs requires dedicated low energy experiments, in contrast to the WIMPs which are suitably searched in the facilities of the high energy frontier such as the CERN Large Hadron Collider (LHC). A number of methods have been proposed and implemented in the recent years, based on lasers, microwave cavities, strong electromagnetic fields or torsion balances~\cite{JaRi2010}. The OSQAR experiment at CERN combines high intensity laser beams and strong magnetic fields to search for WISPs at this low energy frontier. One of its setups uses the ``Light Shining Through a Wall'' (LSW) method for the search of the WISPs, as considered in a  pioneering work which excluded ALPs with a di-photon coupling constant $g_{\text{A}\gamma\gamma}$ larger than $6.7 \cdot 10^{-7}$ GeV$^{-1}$ for ALPs masses below $10^{-3}$ eV~\cite{Cameron1993}. These exclusion limits were extended in later LSW experiments, which now excludes ALPs with a  di-photon coupling constant $g_{\text{A}\gamma\gamma}$ larger than $6.5 \cdot 10^{-8}$ GeV$^{-1}$ in the massless limit~\cite{Ehret2010}. We present here, following our previous work~\cite{Pugnat2014}, the analysis of the latest dataset taken in 2013, including advanced techniques for the data treatments. As a result we will tighten the current exclusion limits for the ALPs di-photon coupling constant down to $g_{\text{A}\gamma\gamma} <  5.7 \cdot 10^{-8}$ GeV$^{-1}$ in the massless limit.

%------------------------------------------------ 
\section{Experimental Setup and Data Taking}
\label{sec:setup}

\begin{wrapfigure}{r}{0.6\textwidth}
\includegraphics[width=0.6\textwidth]{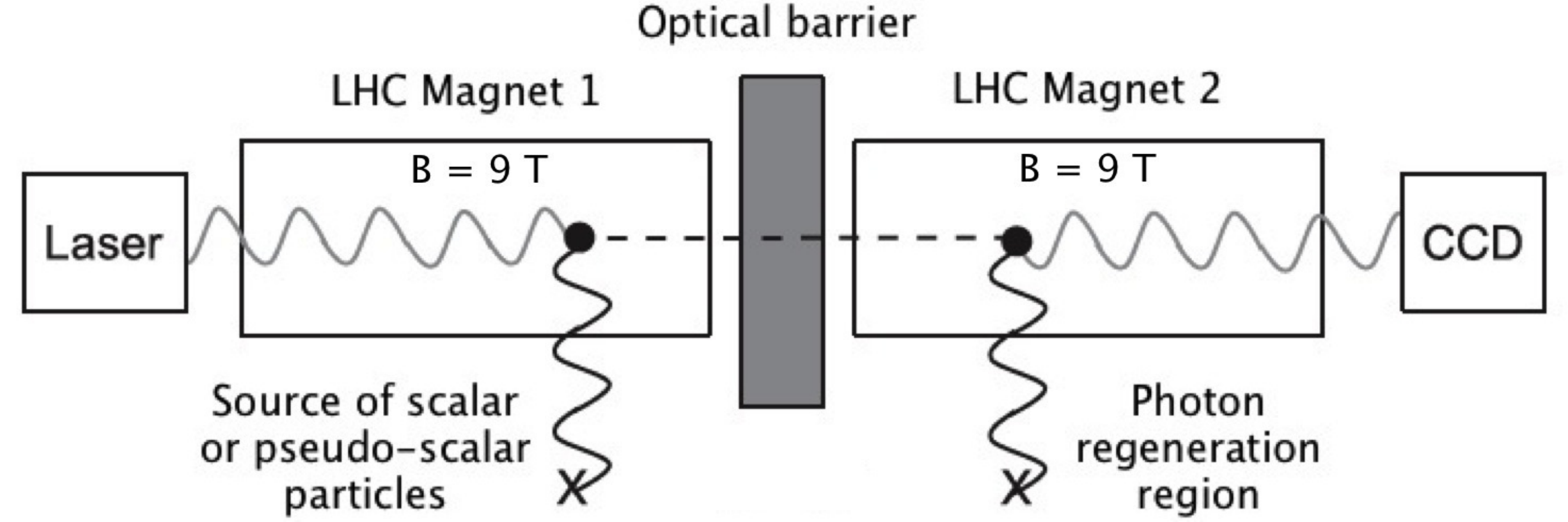}
\caption{Principle of a LSW experiment for ALPs search.}
\label{fig:lsw}
\end{wrapfigure}
LSW experiments are based on the combination of photon-to-WISP and WISP-to-photon double quantum oscillations and interaction weakness of the WISPs with the fermions of the SM. 
The method in the case of ALPs takes advantage of the ALPs di-photon coupling to create ALPs from polarized photons traversing a magnetic field, which will propagate across the optical barrier where the photons are blocked (see Fig.~\ref{fig:lsw}). A magnetic field of same strength is applied in the regeneration domain where the opposite process occurs meaning the ALP produces a photon that is subsequently detected by a CCD.
The Lagrangian density of the interaction of a pseudoscalar ALPs field $\mathcal{A}$, such as the axion field, with the electromagnetic field $F_{\mu\nu}$ has the generic form $\mathcal{L}_{int} = -\frac{1}{4} g_{\text{A}\gamma\gamma}~\mathcal{A}~F_{\mu\nu}\widetilde{F}^{\mu\nu} = g_{\text{A}\gamma\gamma}~\mathcal{A}~\mathbf{E} \cdot \mathbf{B}$, where $\widetilde{F}^{\mu\nu} = \frac{1}{2}\epsilon^{\mu\nu\alpha\beta}F_{\alpha\beta}$ is the dual of $F_{\mu\nu}$ and $g_{\text{A}\gamma\gamma}$ the ALP di-photon coupling constant. This points out that the photons from the incoming laser beam must be polarized parallel to the magnetic field for optimum conversion. With a scalar ALPs field $\mathcal{A}$ the interaction takes the generic form $\mathcal{L}_{int} = -\frac{1}{4} g_{\text{A}\gamma\gamma}~\mathcal{A}~F_{\mu\nu} F^{\mu\nu} = g_{\text{A}\gamma\gamma}~\mathcal{A}~\frac{1}{2}(\mathbf{E}^2 - \mathbf{B}^2)$, in which case the photons from the incoming laser beam must be polarized perpendicular to the magnetic field for optimum conversion. 
In all cases the probability of an ALP-to-photon ($A \rightarrow \gamma$) or of a photon-to-ALP ($\gamma \rightarrow A$) conversion is given by~\cite{Sikivie1983}:
\begin{equation}
P_{\gamma\leftrightarrow A} = \frac{1}{4} {(g_{\text{A}\gamma\gamma} BL)}^2
        {\left(
            \frac{2}{qL} \sin \frac{qL}{2}
        \right)}^2
    \label{eq:probability}
\end{equation}
in units of Heaviside-Lorentz system $(\hbar=c=1)$. $q = | k_\gamma - k_A |$ stands for the momentum transfer, where $k_\gamma = \omega$ is the momentum of the photon of energy $\omega$ and $k_A =
\sqrt{(\omega^2 - m_A^2)}$ the momentum of the ALP of mass $m_A$. The concept of LSW experiments inherits the probability given in Eq.~(\ref{eq:probability}) twice so the overall probability for photon regeneration is $P_{\gamma \rightarrow A \rightarrow \gamma} = {(P_{\gamma\leftrightarrow A})}^2$. This one being proportional to the $4-th$ power of ${(BL)}$ stresses the necessity of the strongest magnetic field $B$ over the longest optical path length $L$. Taking $\eta$ as the photon detection efficiency and $P$ as the optical power, the flux of detected reconverted photons is then given by 
\begin{equation}
    \frac{dN}{dt} = \frac{P}{\omega} \eta \; {(P_{\gamma\leftrightarrow W})}^2
    \label{eq:flux}
\end{equation}

The experimental setup of the OSQAR photon regeneration consists of two LHC dipole magnets separated by an optical barrier as schematized in Fig.~\ref{fig:lsw}. Each magnet is cooled down to 1.9\,K with superfluid He and provides an uniform transverse magnetic field with a strength of 9\,T over a magnetic length of 14.3\,m. A diode-pumped solid-state laser from Coherent, Inc. has been used to deliver 15\,W of optical power at a single wavelength of 532\,nm (2.33\,eV). The photon beam is linearly polarized with a vertical orientation parallel to the magnetic field, which is  suited for the search of axions and pseudoscalar ALPs. A $\lambda/2$ wave-plate with antireflective coating layers oriented at $45^\circ$ was inserted at the laser exit to align the polarization in the horizontal direction for the search of scalar ALPs. The laser light after the second magnet was focussed by an optical lens on a Spectrum One $LN_2-$cooled CCD detector from Instrument SA, Inc. The CCD chip is composed by a 2D array of $1152 \times 298$ square pixels with a $26\,\mu m$ size. Dark current and readout noise are given as 1 - 3 e$^-$/pixel/hour and 4 - 10 e$^-$ rms/pixel. The overall quantum efficiency, including the gain factor, is measured to $33.7\pm 0.7\%$. 

The data-taking was performed in August 2013. Each run consisted typically of nine \textit{frames} of recorded signal distribution on the CCD. The first three frames are taken during one minute with strongly attenuated laser power and without the optical barrier to record the laser position on the CCD, which  defines the expected signal region and ensures the alignment of the experimental setup. The laser power then is set to maximum and the optical barrier positioned for the actual measurements. To receive as high as possible signal to noise ratio, with acceptable risk of cosmic rays contamination of the region of interest, three consecutive frames were recorded with exposure time of 2700\,s for each. The run ends by removing the optical barrier and recording again the laser position on the CCD in three independent frames of one minute. Considering conservatively only runs with very stable laser position on the CCD the whole collected data amounts to integrated measuring time of 19.5\,h for pseudoscalar ALPs search and 26.25\,h for scalar ALPs search. Background frames were also recorded with laser beam switched off.

%------------------------------------------------ 
\section{Data Analysis}
\label{sec:analysis}

The laser-beam was focused on one pixel of the CCD. It however might show a tiny displacement, owing to thermal or mechanical noise around the experimental installation. This was scrutinized to avoid enlarging the area of the possible reconverted photons on the CCD, which degrades the precision  of the data analysis. No displacement was observed during the three consecutive one minute recorded laser positions either before or after signal recording. Hence the signal region is defined per run by the recorded laser positions before and after the signal exposure. If no displacement is observed then the signal region is exactly one pixel wide. If the laser-position before and after the signal recording are apart then the sensitive signal area is determined by a linear interpolation in time. The runs with a  displacement exceeding 2 pixels are discarded.

\begin{figure}[h!]
    \centering
    \includegraphics[width=0.49\textwidth]{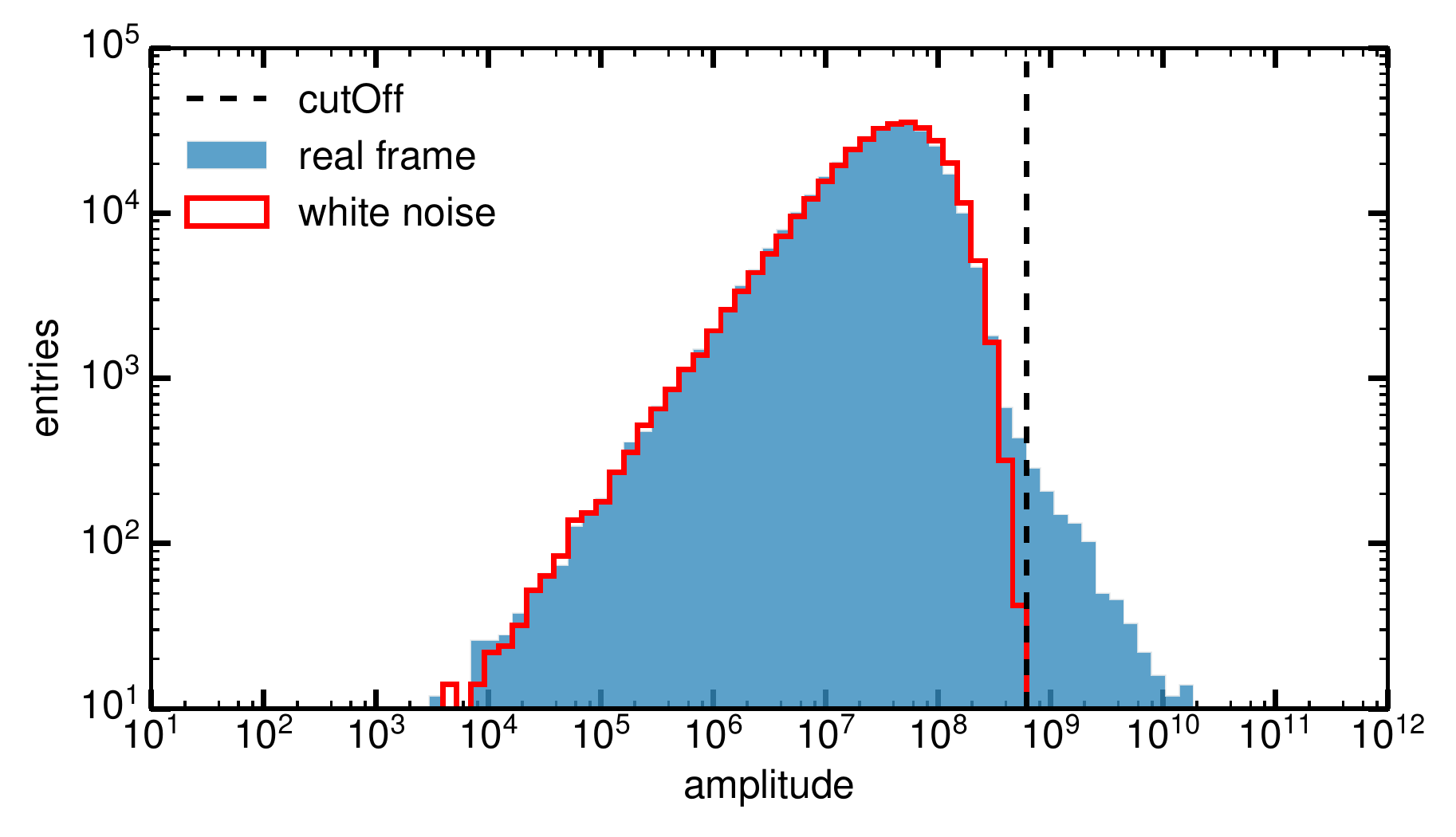}
    \includegraphics[width=0.49\textwidth]{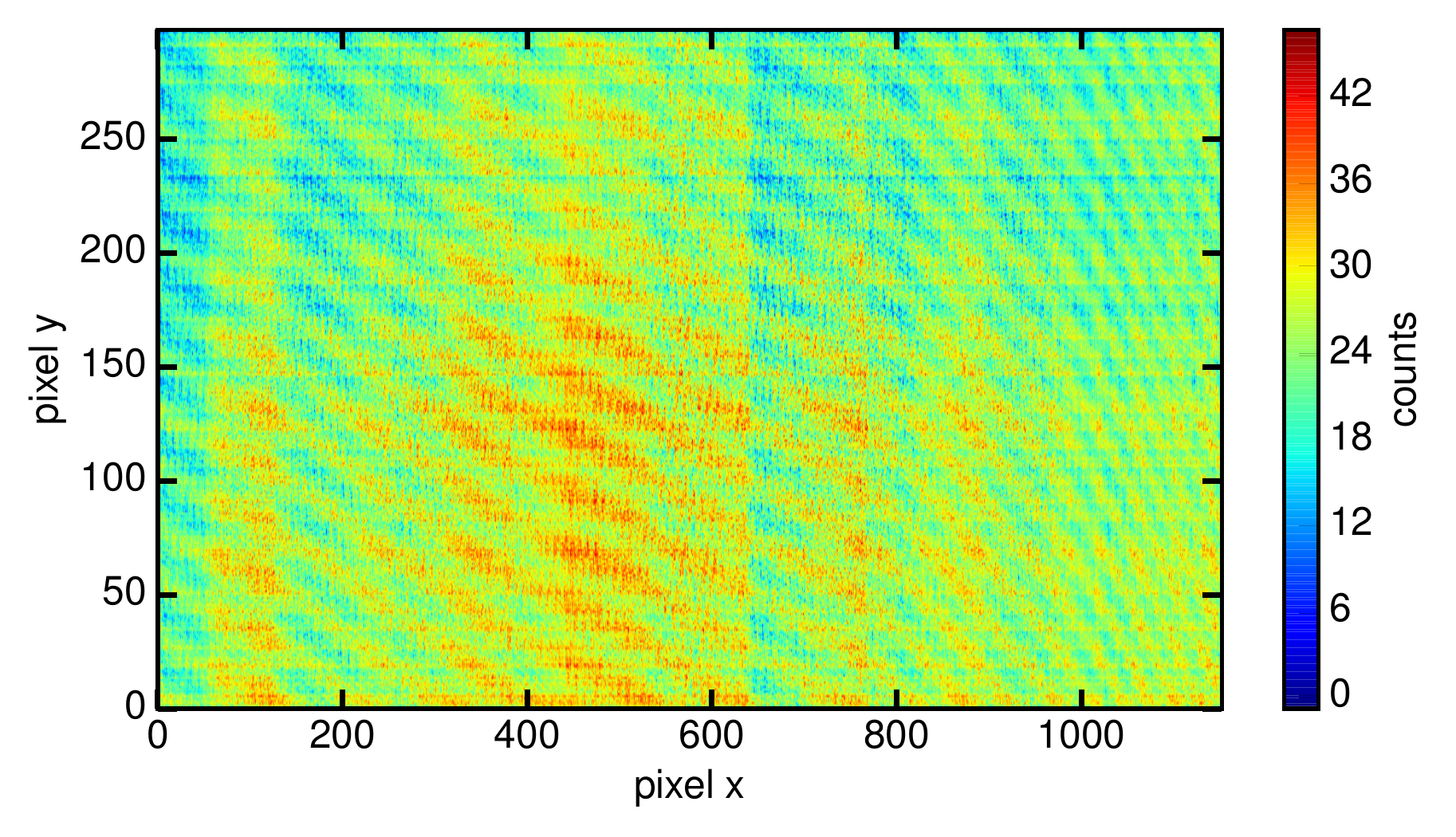}
    \caption{Left: Comparison of the FFT amplitude spectrum of a real frame (blue) with gaussian white noise (red). Right: Inverse transform into spatial domain of the subset $\Omega$ (see text).}
\label{fig:FFTfiltering}
\end{figure}

The basic idea of the data analysis is to compare the recorded photon counts in the signal region to the recorded photon counts outside this region, i.e.\ the background region not exhibited to possible ALPs signals. The sensitivity of the experiment is determined by the width of the distribution of these background counts, which thus must be first cleaned from contamination 
\begin{wrapfigure}{r}{0.5\textwidth}
\includegraphics[width=0.5\textwidth]{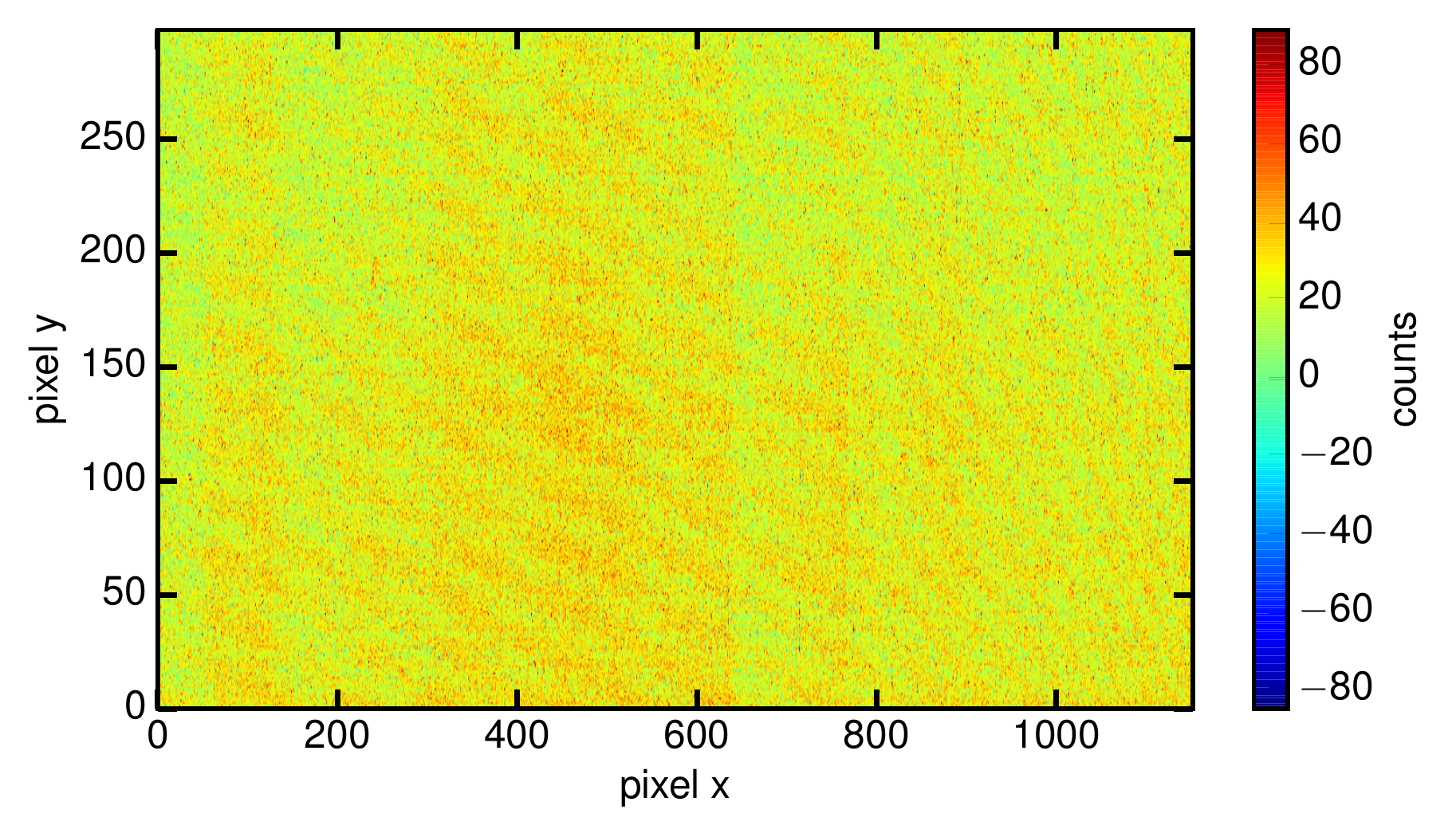}
\caption{Example frame after removal of the contamination by cosmic ray events. A wavelike structure spread over whole chip reveals itself.}
\label{fig:contaminated}
\end{wrapfigure}
by cosmic ray events and corrected for possible defect of flatness. An impact of cosmic ray reveals itself as a series of contiguous pixels with photon counts exceeding a certain threshold in terms of standard deviation. Cosmic ray events are removed in each data frame separately. A median filtered copy is calculated, with fixed median-filter kernel size of $8\times8$ pixels, for the purpose of replacing areas of hot pixels in the original data frame with its median filtered correspondents. The signal region is explicitly excluded from this procedure in order to preserve possible ALPs signatures. 
The recorded frames also show a wavelike structure evolving over the whole CCD chip, as displayed in Fig.~\ref{fig:contaminated}, which was confirmed to be inherent to the used CCD detector by recorded  frames without laser beam. It was important to correct this lack of flatness since it decreases the sensitivity of the experiment by broadening the width of the distribution of the background counts. A filter technique based on Fast-Fourier-Transformations (FFT) has been applied to each data frame individually. 
The two dimensional FFT of the frame is created and the power spectrum $|A(f_x,f_y)|^2$ is calculated\footnote{It is pointed out that the frequency spectrum of a real valued distribution is point symmetric to the origin frequency $A(0,0)$, which corresponds to the sum of all pixels in spatial domain. $A(0,0)$ may represent the average depending on the normalization convention of the FFT.}.
A subset $\Omega \subset\{\,A(f_x, f_y)\,\}$ of the frequency spectrum, subject to cause the wavelike structure in the spatial domain is identified by comparing the amplitudes with an artificial amplitude spectrum of pure gaussian white noise with same mean and standard deviation (Fig.~\ref{fig:FFTfiltering}~Left). 
$\Omega = \{A(f_x,f_y) \;\bigl|\bigr.\; |A(f_x, f_y)|^2 > \text{cutOff} \;\}$ where \textit{cutOff} is defined as the highest amplitude occurring in the white noise spectrum and is individually determined for each frame. The frequency set $\Omega$ is then transformed back into spatial domain exposing the shape of the wavelike structure (Fig.~\ref{fig:FFTfiltering}~Right) and is subtracted from the original data frame.
Since the origin frequency $A(0,0)$ is per definition contained in $\Omega$, this procedure also results in the normalization of the data frames in terms of setting their average pixel count to zero. To ensure that no possible ALPs signal will be impacted by this procedure, an artificial $2\sigma$ \textit{fake signal} imposed to the signal region of a dedicated copy of the frame is checked to be still present after the filtering.

\begin{wrapfigure}{r}{0.5\textwidth}
\includegraphics[width=0.5\textwidth]{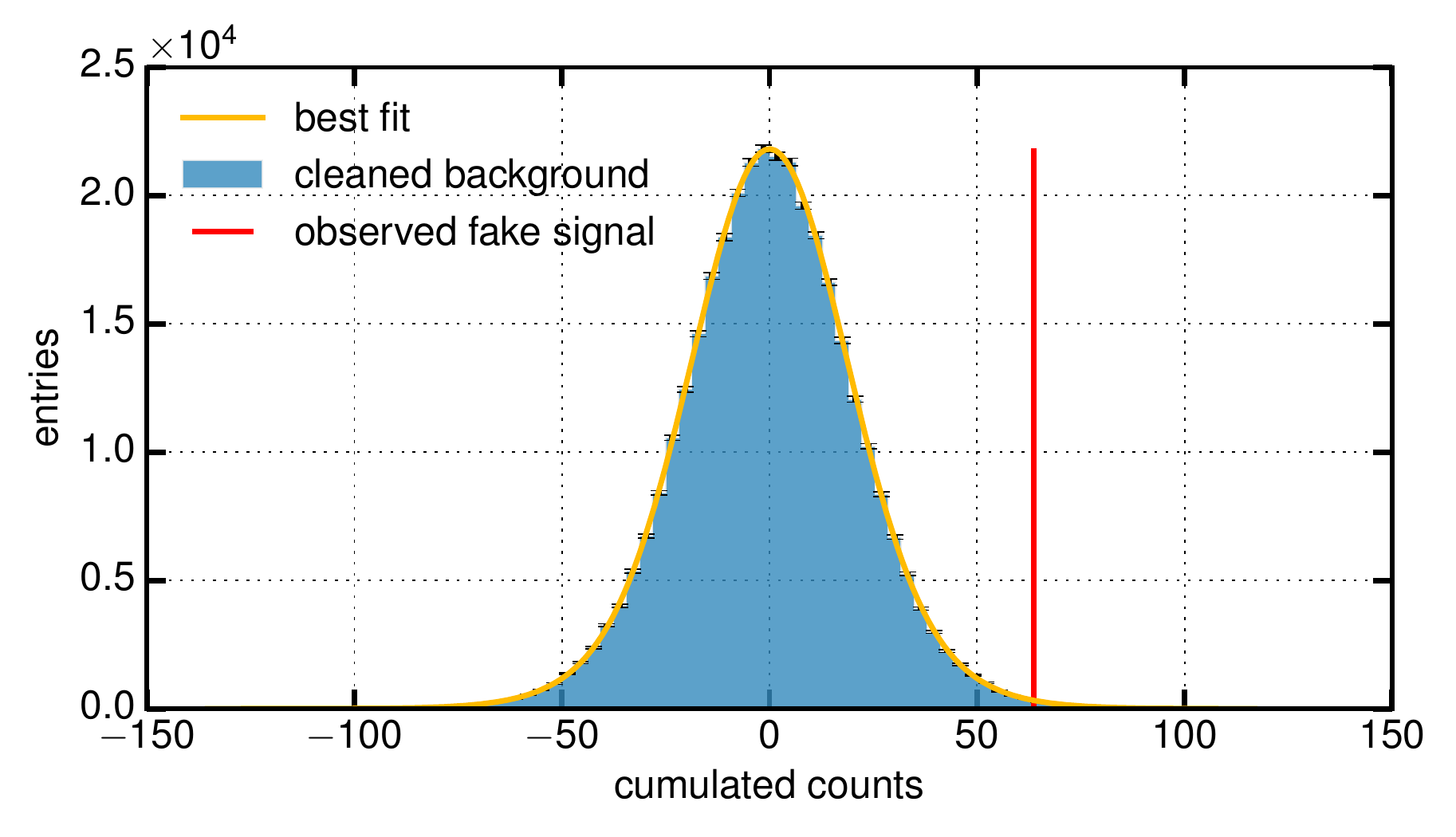}
\caption{An example of background distribution obtained after cosmic events removal and defect of flatness correction (blue) and  signal counts (red) of beforehand imposed fake signal.}
\label{fig:fakeSig}
\end{wrapfigure}
The whole treatment, i.e.\ cosmic events removal and defect of flatness correction, was tested on the background frames to which an artificial signal was imposed corresponding to $m_A = 10^{-4}$ eV and $g_{\text{A}\gamma\gamma} = 8.5 \cdot 10^{-8}$ GeV$^{-1}$. It has then been applied to all data-frames, corresponding to a specific laser polarization. 
The distribution of the recorded photon count in an example background region and also the imposed signal are shown in Fig.~\ref{fig:fakeSig}.
The counts in the signal regions are evaluated and the resulting background distributions are fitted by the sum of two gaussian functions with a shared parameter for the mean value. Typical
$\chi^2/\text{n.d.f}$ values resulting from the fitting procedure are in the range of 0.8 to 1.5 documenting a good parametric description.

%------------------------------------------------ 
\section{Exclusion Limits}

\begin{figure}[h!]
    \centering
    \includegraphics[width=0.49\textwidth]{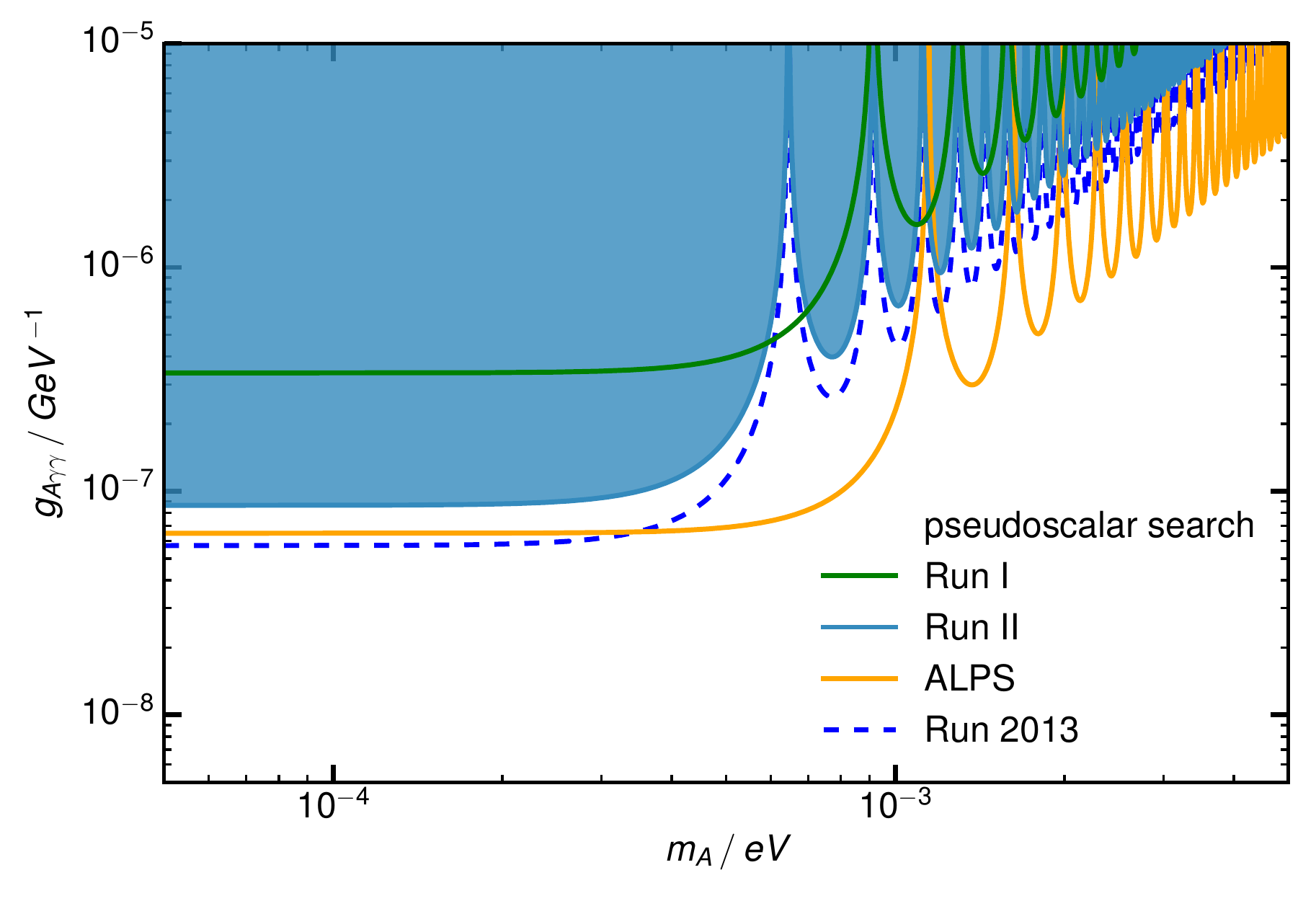}
    \includegraphics[width=0.49\textwidth]{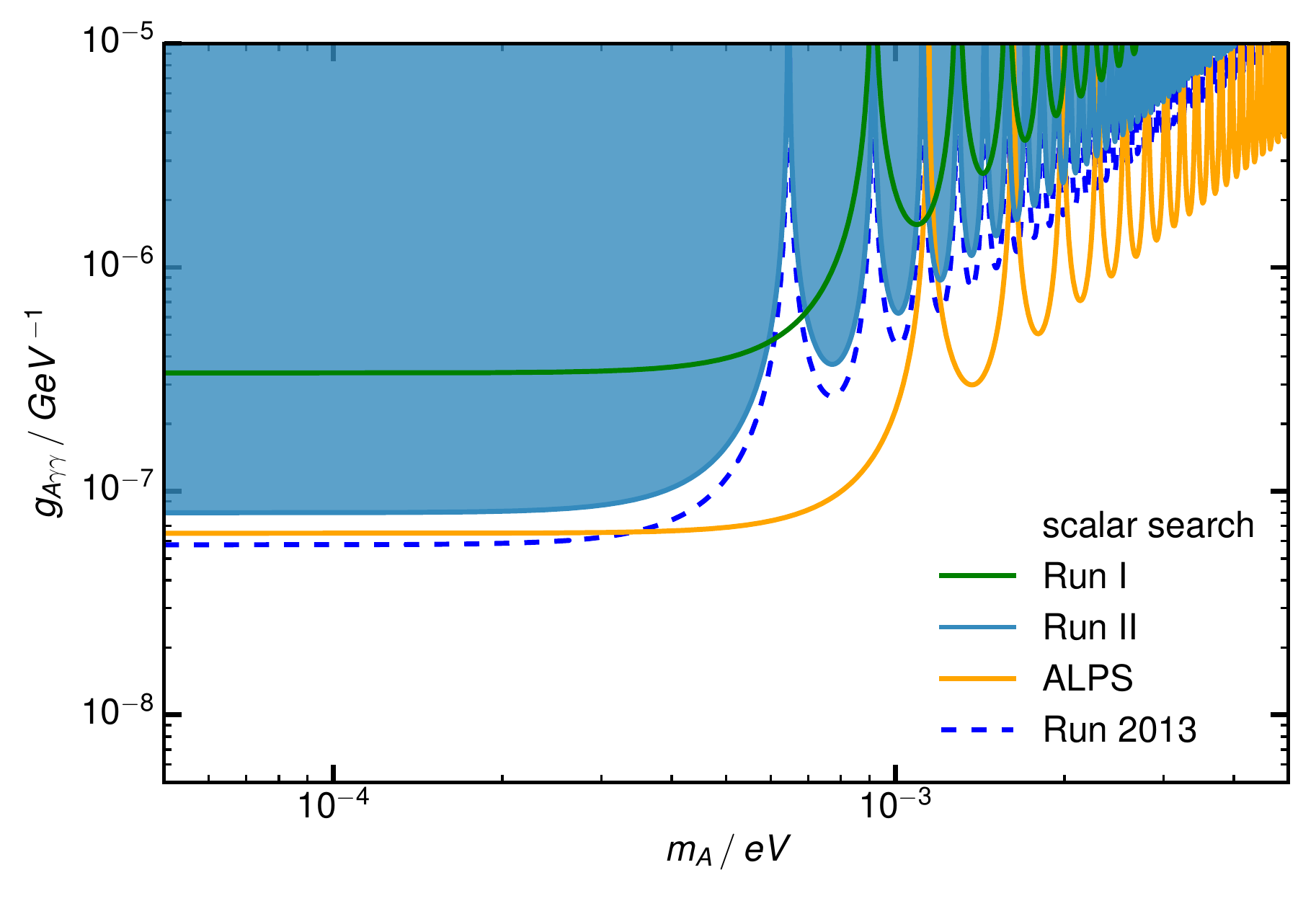}
    \caption{Exclusion limits inferred from the pseudoscalar and scalar search in vacuum for the present 
     experiment compared to previous runs and the latest ALPS result in vacuum~\cite{Ehret2010}.}
\label{fig:limit}
\end{figure}

No excess of counts is detected  in the signal region whatever the polarization of the photons, parallel or perpendicular to the magnetic field.
Exclusion limits on the ALP di-photon coupling strength $g_{\text{A}\gamma\gamma}$ and the ALP mass $m_A$ were then derived via Eq.~(\ref{eq:probability}) \& Eq.~(\ref{eq:flux}).
The 95\% confidence limits (C.L.) are based on the Bayesian inference with the likelihood function $\mathcal{B} =  \prod_{i} \mathcal{B}_i (n_{obs;i}| \frac{dN}{dt})  \cdot \pi(\frac{dN}{dt})$ and a flat prior $\pi(dN/dt)$ for the signal parameter.
The index $i$ runs over the integer number of runs corresponding to a certain polarization state and $\mathcal{B}_i (n_{obs;i}| \frac{dN}{dt})$ represents the background parametrization of the $i$-th frame including an additional poissonian signal contribution of expectation value $dN/dt$ times the frames
exposure time.
\begin{wraptable}{r}{0.65\textwidth}
\centerline{\begin{tabular}{lc|c|c}
\hline
                &                 & \textsc{Pseudoscalar}                 &                 \textsc{Scalar} \\
\hline
$dN_{95\% C.L.}/dt$   & [Hz]   &  $2.07 \cdot 10^{-3}$   &  $2.14 \cdot 10^{-3}$ \\
$g_{\text{A}\gamma\gamma}^{m_A \rightarrow 0}$   &  [Gev$^{-1}$]     &    $5.71 \cdot 10^{-8}$   & $5.76 \cdot 10^{-8}$ \\
\hline
\end{tabular}}
\caption{$95\% \text{C.L.}$ limit on flux of reconverted photons and resulting limit on di-photon coupling constant in massless limit.}
\label{tab:limits}
\end{wraptable}
From the posterior distribution for the signal parameter $dN/dt$ the $95\%\text{C.L.}$ limit on the reconverted photon flux can be set. Table~\ref{tab:limits} summarizes the obtained values and the resulting limit on $g_{\text{A}\gamma\gamma}$ for a vanishing ALP mass ($m_A\rightarrow 0$).
The functional relation of Eq.~(\ref{eq:probability}) leads to limits of $g_{\text{A}\gamma\gamma}$ in dependence of the ALP mass $m_A$ as illustrated in Fig.~\ref{fig:limit} for the scalar- and pseudoscalar search, where also results of the previous experiments are shown.

%------------------------------------------------ 
\section{Conclusion}
\label{sec:conclusion}
We have reported on the latest results obtained from a conservative analysis of the OSQAR LSW experiment for ALPs search conducted in the year 2013. No signal was detected. The 95\% derived confidence limits exclude di-photon coupling strength above $5.7 \cdot 10^{-8}$ GeV$^{-1}$ in the massless limit, giving the so far most stringest constraints on ALPs in LSW-type experiments.

\section{Bibliography}

% ****************************************************************************
% BIBLIOGRAPHY AREA
% ****************************************************************************

\begin{footnotesize}

\end{footnotesize}

% ****************************************************************************
% END OF BIBLIOGRAPHY AREA
% ****************************************************************************

\end{document}